 \documentclass[final,5p,times,twocolumn]{elsarticle}

\usepackage{amssymb}
\journal{Physica C}

\begin{document}

\begin{frontmatter}

%% Title, authors and addresses

%% use the tnoteref command within \title for footnotes;
%% use the tnotetext command for theassociated footnote;
%% use the fnref command within \author or \address for footnotes;
%% use the fntext command for theassociated footnote;
%% use the corref command within \author for corresponding author footnotes;
%% use the cortext command for theassociated footnote;
%% use the ead command for the email address,
%% and the form \ead[url] for the home page:
%% \title{Title\tnoteref{label1}}
%% \tnotetext[label1]{}
%% \author{Name\corref{cor1}\fnref{label2}}
%% \ead{email address}
%% \ead[url]{home page}
%% \fntext[label2]{}
%% \cortext[cor1]{}
%% \address{Address\fnref{label3}}
%% \fntext[label3]{}

%Corresponding author \\

%Hidekazu Mukuda \\

%Address: 1-3, Machikaneyama-cho, Toyonaka, Osaka 560-8531, Japan \\

%TEL: +81-6-6850-6437 \\

%FAX: +81-6-6850-6437 \\

%Email: mukuda@mp.es.osaka-u.ac.jp  \\

%\clearpage

\title{Novel Superconducting Characteristics and Unusual Normal-State Properties in Iron-based Pnictide Superconductors: 
 $^{57}$Fe-NMR and $^{75}$As-NQR/NMR studies in REFeAsO$_{1-y}$ (RE = La, Pr, Nd) and Ba$_{0.6}$K$_{0.4}$Fe$_2$As$_2$}

%% use optional labels to link authors explicitly to addresses:
%% \author[label1,label2]{}
%% \address[label1]{}
%% \address[label2]{}

\author[1,3]{H. Mukuda}
\author[1]{N. Terasaki}
\author[1,3]{M. Yashima} 
\author[1]{H. Nishimura} 
\author[1]{Y. Kitaoka} 
\author[2,3]{A. Iyo} 

\address[1]{Graduate School of Engineering Science, Osaka University, Toyonaka, Osaka 560-8531, Japan }
\address[2]{National Institute of Advanced Industrial Science and Technology (AIST), Umezono, Tsukuba 305-8568, Japan }
\address[3]{JST-TRIP, 1-2-1 Sengen, Tsukuba 305-0047, Japan}

\begin{abstract}

We discuss the novel superconducting characteristics and unusual normal-state properties of iron (Fe)-based pnictide superconductors REFeAsO$_{1-y}$ (RE=La,Pr,Nd) and Ba$_{0.6}$K$_{0.4}$Fe$_2$As$_2$($T_{c}=$ 38 K) by means of $^{57}$Fe-NMR and $^{75}$As-NQR/NMR. In the superconducting state of LaFeAsO$_{0.7}$ ($T_{c}=$ 28 K), the spin component of the $^{57}$Fe-Knight shift decreases to almost zero at low temperatures, which provide firm evidence of the superconducting state formed by spin-singlet Cooper pairing. The nuclear spin-lattice relaxation rates $(1/T_{1})$ in LaFeAsO$_{0.7}$ and Ba$_{0.6}$K$_{0.4}$Fe$_2$As$_2$ exhibit a $T^{3}$-like dependence without a coherence peak just below $T_{c}$, indicating that an unconventional superconducting state is   commonly realized in these Fe-based pnictide compounds. 
All these events below $T_c$ are consistently argued in terms of an extended s$_{\pm}$-wave pairing with a sign reversal of the order parameter among Fermi surfaces. In the normal state, $1/T_1T$ decreases remarkably upon cooling for both the Fe and As sites of LaFeAsO$_{0.7}$. In contrast, it gradually  increases upon cooling in Ba$_{0.6}$K$_{0.4}$Fe$_2$As$_2$.
Despite the similarity between the superconducting properties of these compounds, a crucial difference was observed in their normal-state properties depending on whether electrons or holes are doped into the FeAs layers. These results may provide some hint to address a possible mechanism of Fe-based pnictide superconductors.

\end{abstract}

\begin{keyword}
%% keywords here, in the form: keyword \sep keyword
superconductivity, iron-based pnictide, LaFeAsO, (Ba,K)Fe$_2$As$_2$, NMR, NQR
%% PACS codes here, in the form: \PACS code \sep code
74.70.-b,74.20.Rp,76.60.-k
%  CeIrSi3_  74.70.Tx,74.25.Dw,74.62.Fj,76.60.-k
%% MSC codes here, in the form: \MSC code \sep code
%% or \MSC[2008] code \sep code (2000 is the default)

\end{keyword}

\end{frontmatter}

%% \linenumbers

%% main text

%%%%%%%%%%%%%%%%%     Introduction     %%%%%%%%%%%%%%%%%%%%%%%%%

\section{Introduction}

The discovery of superconductivity (SC) in the iron (Fe)-based oxypnictide  LaFeAsO$_{1-x}$F$_x$, reaching a superconducting transition temperature $T_c$ = 26 K, has attracted considerable interest in the fields of condensed-matter physics and material science \cite{Kamihara2008}.
Shortly after this discovery, it was reported that $T_c$ of LaFeAsO$_{0.89}$F$_{0.11}$ increases up to 43 K upon the application of pressure \cite{Takahashi}, and the replacement of the La site by other rare earth (RE) elements significantly increases $T_c$ up to more than 50 K \cite{Ren1,Kito,Ren2,GdFeAsO}. These findings have provided a new material base for searching high-$T_c$ SC. The structure of mother materials contains an alternate stacking of RE$_2$O$_2$ and Fe$_2$As$_2$ layers along the c-axis where the Fe atoms of the FeAs layer are located in a fourfold coordination forming a FeAs$_{4}$ tetrahedron. The mother material LaFeAsO is a semimetal with a stripe antiferromagnetic (AFM) order with ${\bf Q}=(0,\pi)$ or $(\pi,0)$ \cite{Cruz}. The substitution of fluorine for oxygen and/or oxygen deficiencies at the LaO layer yield a novel SC \cite{Kamihara2008,Ren1,Kito,Ren2,GdFeAsO}. In particular, a very sharp superconducting transition in resistance under $P$ for NdFeAsO$_{0.6}$ ensures a homogeneous electronic state even in an oxygen-deficient sample \cite{Takeshita}.  Remarkably, Lee et al. found that $T_{c}$ increases to the maximum value of 54 K when the FeAs$_{4}$ tetrahedron is transformed into a regular one \cite{C.H.Lee}.  

Another family of FeAs-based superconductors without oxygens has been reported in the ternary compound Ba$_{1-x}$K$_{x}$Fe$_{2}$As$_{2}$ with $T_c=$ 38 K \cite{Rotter}. In this compound, layers consisting of edge-sharing FeAs$_{4}$ tetrahedra are separated by Ba(K) layers.  Moreover, SC was also reported in $\alpha$-FeSe with $T_c=$ 8 K \cite{Hsu}. This compound is composed of stacking layers of FeSe, resembling the FeAs layers in LaFeAsO$_{1-x}$F$_{x}$, but containing neither any Ba(K) atoms nor LaO sheets.  The present experimental facts suggest that systematic studies of the local electronic state at the Fe site are quite important to elucidate the origin of SC in the iron-based compounds.

In this paper, we report $^{57}$Fe-NMR and $^{75}$As-NQR/NMR studies of the superconducting and normal-state properties of Fe-based pnictide high-$T_c$ superconductors.  We review the previous results on REFeAsO$_{1-y}$ published elsewhere \cite{Mukuda,Terasaki}, and present the recent results on Ba$_{0.6}$K$_{0.4}$Fe$_2$As$_2$.

%%%%%%%%%%%%%%%%%%%%%    Experimental    %%%%%%%%%%%%%%%%%%%%%%%%%%%%%

\section{Experimental}

Polycrystalline samples of LaFeAsO$_{1-y}$, PrFeAsO$_{0.6}$, NdFeAsO$_{0.6}$, and Ba$_{0.6}$K$_{0.4}$Fe$_2$As$_2$ were synthesized via the high-pressure synthesis technique described elsewhere \cite{Kito}. 
Although the real oxygen content of the samples may be greater than the nominal (intended) values due to the oxidation of the starting RE elements, powder X-ray diffraction measurements indicate that these samples are almost entirely composed of a single phase. 
The $T_c$s for all samples were determined by susceptibility measurement, which indicated a marked decrease due to the onset of SC below $T_c=$ 20 K, 28 K, 28 K, 22 K, 46 K, and 53 K for LaFeAsO$_{0.75}$, $^{57}$Fe-enriched LaFeAsO$_{0.7}$, LaFeAsO$_{0.6}$,  LaFeAsO$_{0.6}\sharp2$, PrFeAsO$_{0.6}$, and NdFeAsO$_{0.6}$, respectively. 
Note that the lattice parameters $a$ = 4.0226 \r{A} and $c$ = 8.7065 \r{A} of  $^{57}$Fe-enriched LaFeAsO$_{0.7}$ are very close to those of LaFeAsO$_{0.6}$ ($a$ = 4.0220 \r{A} and $c$ = 8.7110 \r{A}), indicating that the physical properties of both samples are compatible. 
$T_c$ of LaFeAsO$_{0.6}\sharp2$ is lower than that of LaFeAsO$_{0.6}$ because the former is in an overdoped regime. This result is corroborated by the fact that the lattice parameters of LaFeAsO$_{0.6}\sharp2$ are smaller than those of LaFeAsO$_{0.6}$ ($T_c=$ 28 K).
The samples were moderately crushed into powder for the NQR/NMR measurements. $^{57}$Fe-NMR  and $^{75}$As-NQR/NMR measurements were performed by using the phase coherent pulsed NMR/NQR spectrometer in the temperature ($T$) range between 4 K and 280 K. $1/T_1$ was measured using the saturation recovery method.

\section{Results and discussion}

\subsection{$^{57}$Fe-NMR study of LaFeAsO$_{1-y}$}

\begin{figure}[tbp]
\begin{center}
\includegraphics[width=0.9\linewidth]{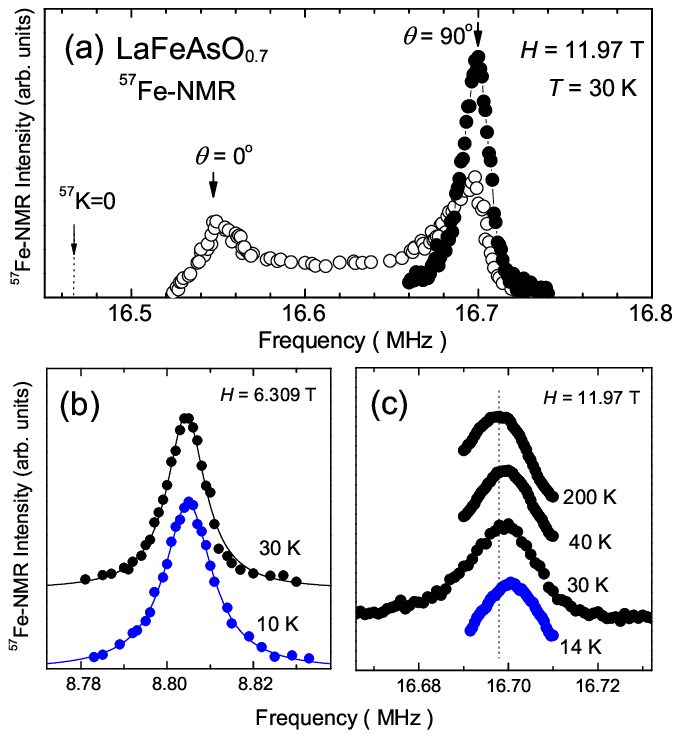}
\end{center}
\caption[]{(color online)
(a) $^{57}$Fe-NMR spectra of LaFeAsO$_{0.7}$ at 30 K and $H$ = 11.97 T in the field parallel ($\bullet$) and perpendicular ($\circ$) to the $ab$-plane. The $T$ dependence of $^{57}$Fe-NMR spectra was obtained at (b) $H$ = 6.309 T and (c) $H$ = 11.97 T parallel to the $ab$-plane($\theta=90^\circ$). 
}
\label{spectrum}
\end{figure}

Figure \ref{spectrum}(a) shows the $^{57}$Fe-NMR spectra for LaFeAsO$_{0.7}$ obtained by a sweeping frequency ($f$) at a magnetic field $H=$ 11.97 T at 30 K. 
For $H$ parallel to the $ab$-plane, a single NMR spectrum is observed with a very narrow linewidth with $\sim$20 kHz. 
This result indicates that the FeAs layers of this sample are rather homogeneous irrespective of the oxygen deficiency in the LaO layer.  For $H$ parallel to the c-axis, respective asymmetric peaks are observed  in the spectra, corresponding to the crystal directions with $\theta=90^\circ$ and $0^\circ$, where $\theta$ is the angle between the field and the c-axis. 
Anisotropic Knight shifts, defined as a shift from $f_{0} =^{57}\gamma_n H$ ($^{57}K=0$), are $^{57}K^\perp \sim 1.413$\% and $^{57}K^\parallel \sim 0.50$\% at 30 K for $\theta=90^\circ$ and $0^\circ$, respectively. 

\begin{figure}[tbp]
\begin{center}
\includegraphics[width=0.9\linewidth]{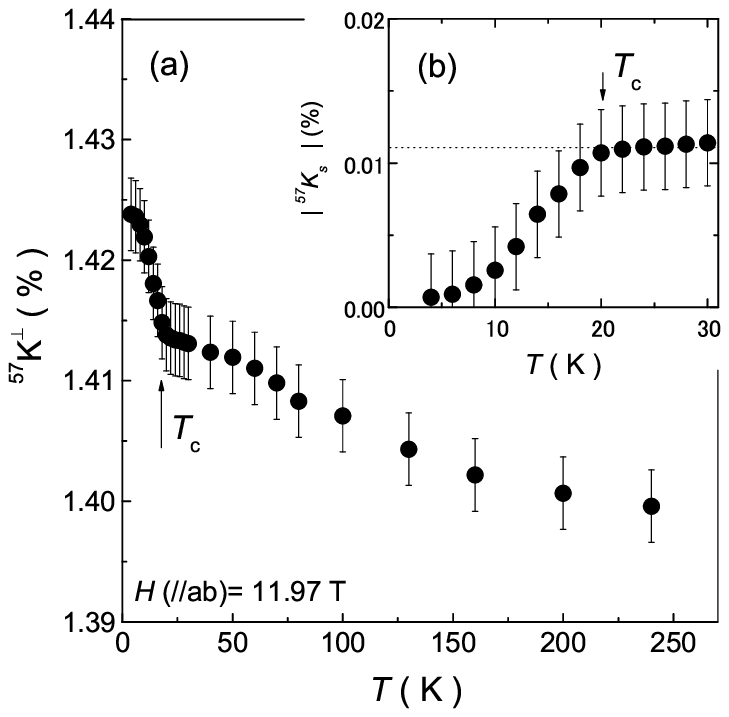}
\end{center}
\caption[]{
(a) $T$ dependence of $^{57}K^\perp$ at $H$ = 11.97 T ($T_{c}(H)\sim$ 20 K) measured by Fe-NMR for LaFeAsO$_{0.7}$. Note that it increases upon cooling, exhibiting a T dependence opposite to those of the $^{75}$As and $^{19}$F sites \cite{Grafe,Ahilan,ImaiJPSJ}. This indicates that the hyperfine-coupling constant $^{57}A_{\rm hf}^\perp$ is negative at the Fe site, originating from the inner core-polarization.
(b) Its spin part $|^{57}K_{\rm s}^\perp|$ evaluated from $|^{57}K^\perp- ^{57}K_{\rm orb}^\perp|$ decreases to zero in the SC state, which provides firm evidence for a spin-singlet Cooper pairing state. 
}
\label{Knightshift}
\end{figure}

Figures~\ref{spectrum}(b) and (c) show the $T$ dependences of the $^{57}$Fe-NMR spectra at $H$= 6.309 T and 11.97 T parallel to the $ab$-plane($\theta=90^\circ$) with $T_{c}(H)\sim$ 24 K and 20 K, respectively.
The $T$ dependence of $^{57}K^\perp$ for $H$ parallel to the $ab$-plane is shown in Fig. \ref{Knightshift}(a). 
The Knight shift comprises the $T$-independent orbital contribution and the $T$-dependent spin contribution, denoted as $^{57}K_{\rm orb}$ and $^{57}K_{\rm s}$, respectively. 
Note that it increases upon cooling, exhibiting a $T$ dependence opposite to those of the $^{75}$As and $^{19}$F sites \cite{Grafe,Ahilan,ImaiJPSJ}. This is because  the hyperfine-coupling constant $^{57}A_{\rm hf}^\perp$  at the Fe site is negative, originating from the inner core-polarization. In this compound, note that $^{57}A_{\rm hf}^\perp=A+4B$, where $A$ is the on-site negative term dominated by the inner core polarization, and $B$ is the transferred positive one from the neighbor Fe sites through direct Fe-Fe and/or indirect Fe-As-Fe bondings. A transferred hyperfine-coupling constant at the As site $^{75}A_{\rm hf}^\perp=4C$ consists of two contributions in the isotropic term of a transferred hyperfine field ($C_{\rm tr}$) and the anisotropic one of a pseudo-dipole field ($C_{\rm dip}$) \cite{Kitagawa}, both of which are induced by neighboring Fe-$d$ spin polarization. 
From a plot of $^{57}K^\perp(T)$ versus $^{75}K^\perp(T)$ with $T$ as an implicit parameter, $^{57}K_{\rm orb}^\perp$ is estimated to be $1.425$\% \cite{Terasaki} using the orbital shift $^{75}K_{\rm orb}^\perp$ at the $^{75}$As site reported in literature \cite{ImaiJPSJ}. Eventually, a spin component of the Knight shift $|^{57}K_{\rm s}^\perp|=|^{57}K^\perp-^{57}K_{\rm orb}^\perp|$, as shown in Fig. \ref{Knightshift}(b), decreases to almost zero well below $T_c$.
This result suggests the possible existence of an isotropic gap in a very-low-temperature regime, providing firm evidence of spin-singlet Cooper pairing through the direct measurement of the local spin susceptibility by means of the $^{57}$Fe-Knight shift . 

\begin{figure}[htbp]
\begin{center}
\includegraphics[width=0.9\linewidth]{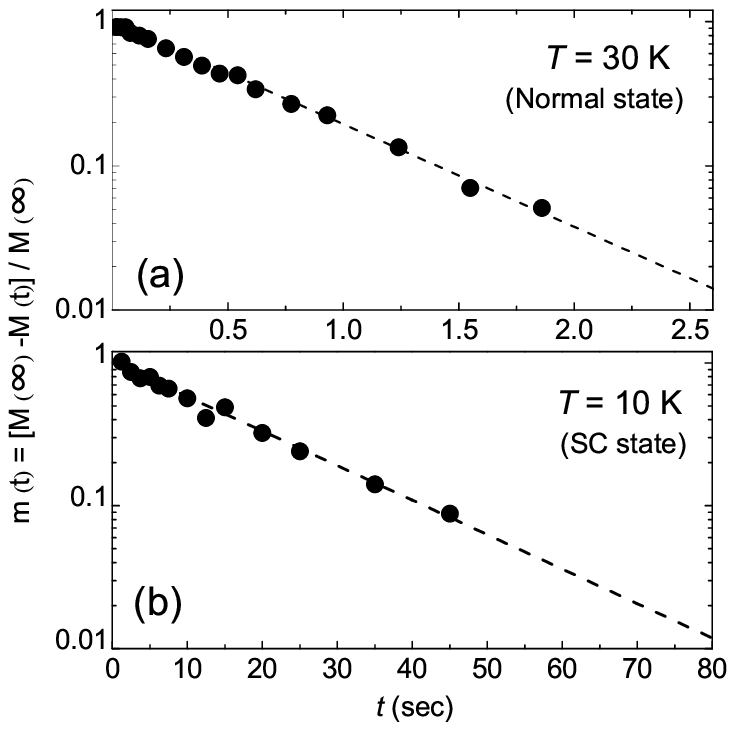}
\end{center}
\caption[]{
The recovery curves of $^{57}$Fe nuclear magnetization under the (a) normal state (30 K)  and (b) SC state (10 K) for LaFeAsO$_{0.7}$ at $f$ = 8.804 MHz and $H$= 6.309 T. In a entire range of $T$, they are uniquely fitted by $^{57}m(t)=(M(\infty)-M(t))/M(\infty)=\exp(-t/T_{1})$ (dashed line).
}
\label{Ferecovery}
\end{figure}

The nuclear spin-lattice relaxation rate $^{57}(1/T_1)$ at the $^{57}$Fe site was determined from a single exponential recovery curve of $^{57}$Fe nuclear magnetization as follows: 
\[
^{57}m(t)\equiv\frac{M(\infty)-M(t)}{M(\infty)}=\exp\left(-\frac{t}{T_{1}}\right),
\]
where $M(\infty)$ and $M(t)$ are the respective nuclear magnetizations for the thermal equilibrium condition and at time $t$ after the saturation pulse. In fact, as shown in Fig. \ref{Ferecovery}, $^{57}T_1$ was uniquely determined from a single exponential function of $^{57}m(t)$ in the entire $T$ range, revealing that the electronic state of the present sample is homogeneous. We have confirmed that $^{57}(1/T_1)$ is isotropic regardless of the crystal direction. 

\begin{figure}[tbp]
\begin{center}
\includegraphics[width=0.9\linewidth]{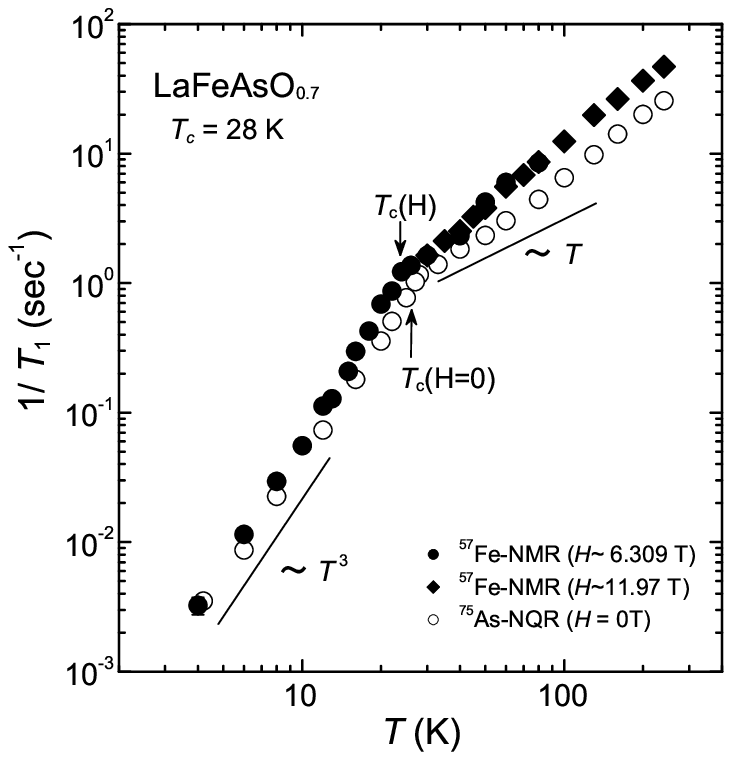}
\end{center}
\caption[]{ $T$ dependence of $^{57}$Fe-NMR $1/T_{1}$ at $H$= 6.309 and 11.97 T for LaFeAsO$_{0.7}$ ($T_{c}=2$8 K), and $^{75}$As-NQR $1/T_{1}$ for LaFeAsO$_{0.6}$ ($T_{c}=28$ K). In the SC state, $1/T_{1}$s at both the Fe and As sites follow a $T^3$-like dependence upon cooling without a coherence peak just below $T_{c}$.
}
\label{fig:T1}
\end{figure}

Figure~\ref{fig:T1} shows the $T$ dependences of $^{57}(1/T_1)$ at $H =$ 6.309 and 11.97 T in the $T$ range of 4$\sim$80 K and 30$\sim$240 K, respectively. In the SC state, $^{57}$Fe-NMR $(1/T_{1})$ exhibits a $T^3$-like dependence without a coherence peak just below $T_{\rm c}(H)=$ 24 K at $H$ = 6.309 T. 
Note that any deviation from the $T^3$ dependence was not observed even at $T/T_{\rm c}=0.17$ well below $T_{\rm c}$.   
Here, it should be noted that in most d-wave superconductors with a line-node gap, such as copper oxides high-$T_c$ superconductors, $1/T_1$ tends to exhibit a $T$ linear dependence at low temperatures, probing the residual density of states (RDOS) at the Fermi level in association with an impurity effect.  

\subsection{$^{75}$As-NQR study of LaFeAsO$_{1-y}$}

\begin{figure}[htbp]
\begin{center}
\includegraphics[width=0.9\linewidth]{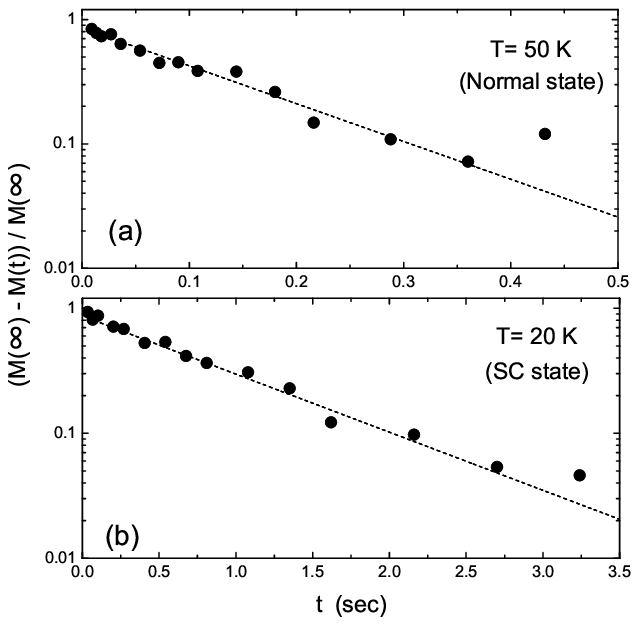}
\end{center}
\caption[]{
The recovery curves of $^{75}$As nuclear magnetization for $^{75}$As-NQR ($H=0$) at (a) the normal state (50 K) and (b) the SC state (20 K) for LaFeAsO$_{0.6}$ at $f=10.05$ MHz. In the entire $T$ range, the recovery curves are uniquely fitted  by $^{75}m(t)_{NQR}=(M(\infty)-M(t))/M(\infty)=\exp(-3t/T_{1})$ (dashed lines).
}
\label{Asrecovery}
\end{figure}

Here, we review the $^{75}$As-NQR $1/T_1$ results for LaFeAsO$_{0.6}$ ($T_{c}=28$ K) at $f=10.05$ MHz and zero field \cite{Mukuda}.  In the $^{75}$As-NQR $T_1$ measurements at $H=0$, the recovery curve of $^{75}$As nuclear magnetization with $I=$ 3/2 is also expressed by a single exponential function as follows: 
\[
^{75}m(t)_{NQR}\equiv\frac{M(\infty)-M(t)}{M(\infty)}=\exp\left(-\frac{3t}{T_{1}}\right).
\]
As shown in Fig. \ref{Asrecovery}, $^{75}m(t)_{NQR}$ was almost fitted by a single exponential function in the SC state and the normal state, ensuring that $1/T_1$ is uniquely determined over the entire $T$ range. 

Figure \ref{fig:T1} shows the $T$ dependence of $^{75}$As-NQR $1/T_{1}$ for LaFeAsO$_{0.6}$ with $T_c$ = 28 K. 
In the SC state, $1/T_{1}$ at the As site exhibits a $T^3$-like dependence without a coherence peak just below $T_{\rm c}=$28 K, resembling the $^{57}$Fe-NMR $1/T_1$ result. 
Considering that $^{57}$Fe-NMR $1/T_1$ measured in the SC mixed state under $H=$ 6.309 T exhibits a $T$ dependence similar to that of $^{75}$As-NQR $1/T_{1}$, it would be expected that the presence of vortex cores would not affect the quasiparticle excitations in the SC state considerably. This implies that the $1/T_1$ measurements can clarify the SC gap structure although it has been measured under $H$. Therefore, the $1/T_{1}\sim T^3$-like behavior with no coherence peak gives firm evidence for an unconventional superconducting nature inherent to the Fe-pnictide superconductors. 
In d-wave superconductors, which also exhibit $1/T_{1}\sim T^3$-like behavior with no coherence peak, a $T_1T=const.$-like behavior was observed at low temperatures, indicating the presence of the RDOS at the Fermi level. This event is well understood in terms of the impurity scattering effect in a unitarity limit in unconventional superconductors with a line-node gap, such as d-wave superconductors. This is not the case in the Fe-based superconductors.  When noting that similar results have been reported in the $^{75}$As-NMR studies on F-substituted LaFeAs(O$_{1-x}$F$_{x}$) \cite{Grafe,Nakai} and PrFeAs(O$_{1-x}$F$_{x}$) \cite{Matano}, and the $^{77}$Se-NMR study on FeSe \cite{Kotegawa} and considering that these materials are far from a clean system, a d-wave model is not suitable for understanding these unconventional $T_1$ behaviors in Fe-based pnictide superconductors.

\subsection{$^{75}$As-NMR study of Ba$_{0.6}$K$_{0.4}$Fe$_2$As$_2$}

\begin{figure}[htbp]
\begin{center}
\includegraphics[width=0.8\linewidth]{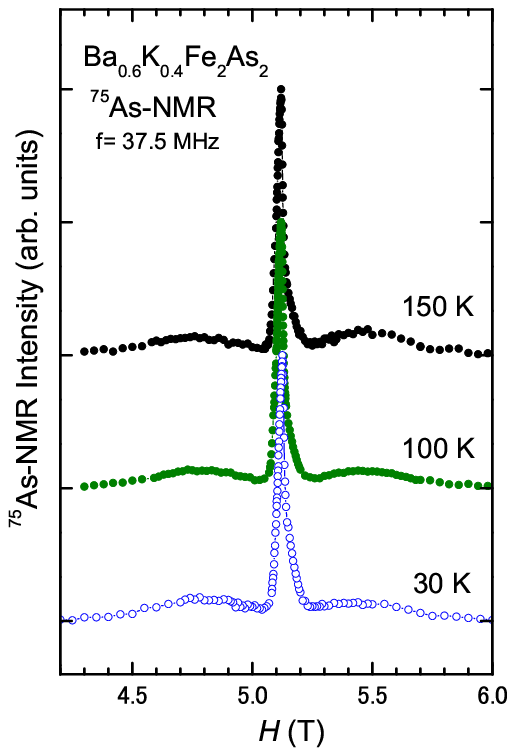}
\end{center}
\caption[]{(color online) $^{75}$As-NMR spectra of Ba$_{0.6}$K$_{0.4}$Fe$_2$As$_2$ at 30 K (SC state), 100 K, and 150 K (normal state). The sharp central peak around $H\sim$ 5.1 T originates from the ($+1/2\leftrightarrow -1/2$) transition and the satellite peaks around $H\sim$ 4.7 and 5.5 T originate from the ($\pm1/2 \leftrightarrow \pm3/2$) transitions. 
}
\label{fig:Ba122spectra}
\end{figure}

Figure \ref{fig:Ba122spectra} shows the typical $^{75}$As-NMR spectra for the oriented powder of Ba$_{0.6}$K$_{0.4}$Fe$_2$As$_2$ with $T_c (H=0)$ = 38 K at $f=$ 37.5 MHz. The sharp central peak observed around $H\sim$ 5.1 T originates from the central transition ($+1/2\leftrightarrow -1/2$) in the $^{75}$As-NMR spectrum. The satellite peaks ($\pm1/2 \leftrightarrow \pm3/2$) around $H\sim$ 4.7 and 5.5 T originate from the first-order perturbation effect of the nuclear quadrupole interaction (NQI), allowing us to estimate the nuclear quadrupole frequency $\nu_Q\sim$ 5($\pm$2) MHz to be larger than 2.2 MHz for the mother compound BaFe$_2$As$_2$ \cite{Fukazawa}. 

\begin{figure}[htbp]
\begin{center}
\includegraphics[width=0.9\linewidth]{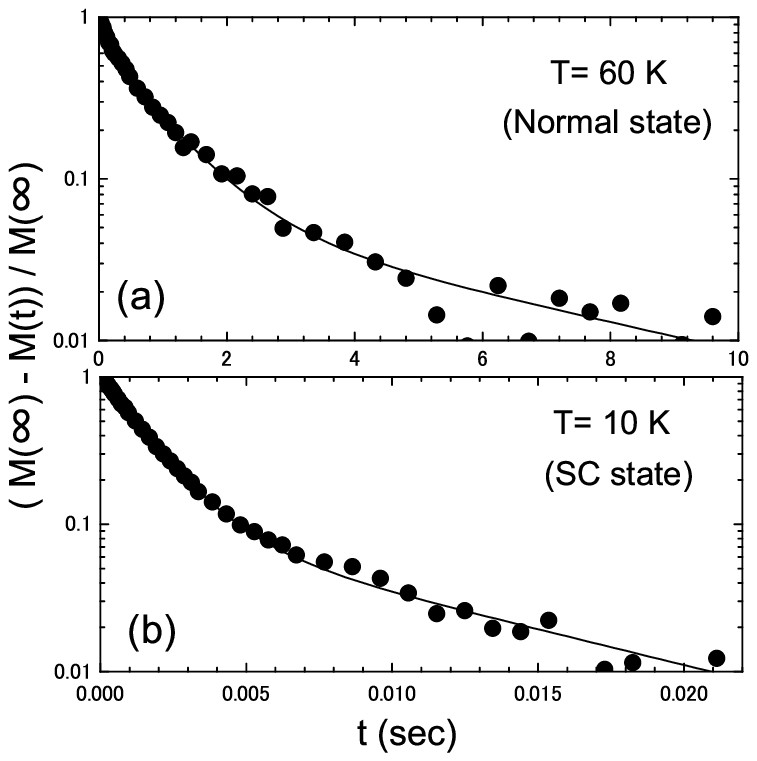}
\end{center}
\caption[]{The recovery curves of $^{75}$As nuclear magnetization for the $^{75}$As-NMR measurement of Ba$_{0.6}$K$_{0.4}$Fe$_2$As$_2$ at (a) the normal state (60 K) and (b) the SC state (10 K), which are uniquely determined  by the theoretical function $^{75}m(t)_{NMR}=(M(\infty)-M(t))/M(\infty)=0.1\exp(-t/T_{1}) + 0.9\exp(-6t/T_{1})$ (solid lines) in the entire $T$ range.
}
\label{fig:Ba122recovery}
\end{figure}

Next, we deal with the SC characteristics probed by the $T_1$ measurement.
The recovery curve of $^{75}$As nuclear magnetization ($I=3/2$) for the $^{75}$As-NMR measurement is expressed by a theoretical curve as follows: 
\[
^{75}m(t)_{NMR}\equiv\frac{M(\infty)-M(t)}{M(\infty)}=0.1\exp\left(-\frac{t}{T_{1}}\right) + 0.9\exp\left(-\frac{6t}{T_{1}}\right).
\]
Figures \ref{fig:Ba122recovery}(a) and (b) show $^{75}m(t)_{NMR}$ in the SC state and the normal state, respectively. 

\begin{figure}[htbp]
\begin{center}
\includegraphics[width=0.8\linewidth]{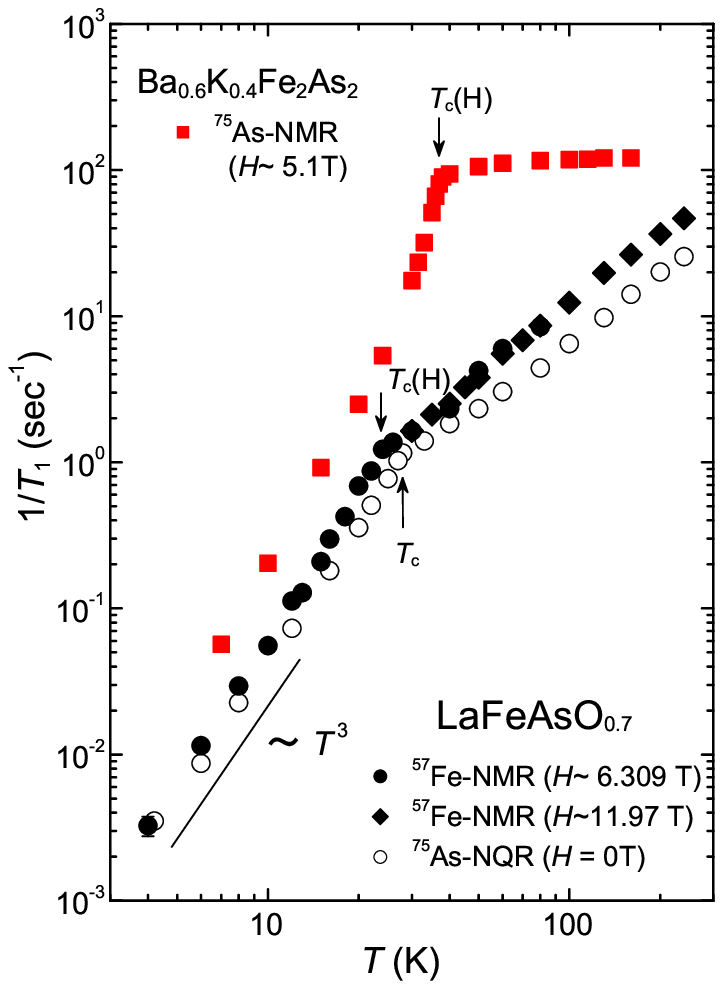}
\end{center}
\caption[]{(color online)
$T$ dependence of $^{75}$As-NMR $1/T_{1}$ at $H\sim$ 5.1 T for Ba$_{0.6}$K$_{0.4}$Fe$_2$As$_2$, along with the results for LaFeAsO$_{0.7}$ ($T_{c}$ = 28 K). In the SC state of Ba$_{0.6}$K$_{0.4}$Fe$_2$As$_2$, $^{75}$As-$1/T_{1}$ drops sharply below $T_{c}(H) =$ 37 K upon cooling without a coherence peak just below $T_c$. 
Although the $T$ dependence of $1/T_{1}$ well below $T_c$ appears to be similar to the $T^3$ behavior, it cannot be exactly reproduced by any simple SC gap model either with line nodes or without nodes. 
It may be related to the characteristics of the multiband SC state observed in Ba$_{0.6}$K$_{0.4}$Fe$_2$As$_2$\cite{ARPES}. 
}
\label{fig:Ba122T1}
\end{figure}

Figure \ref{fig:Ba122T1} shows the $T$ dependence of $^{75}$As-NMR $1/T_{1}$ for Ba$_{0.6}$K$_{0.4}$Fe$_2$As$_2$ at $H\sim$ 5.1 T along with the $1/T_1$ data of LaFeAsO$_{0.7}$.  We note that the present data on Ba$_{0.6}$K$_{0.4}$Fe$_2$As$_2$ are consistent with the result reported by another group \cite{FukazawaSC}. 
In the SC state, $1/T_{1}$ decreases sharply below $T_{c}(H)=$ 37 K upon cooling without a coherence peak just below $T_c$, strongly suggesting an unconventional SC nature. 
Furthermore, the $1/T_{1}$ seems to be close to a $T^3$ dependence well below $T_c$. 
However, we note that the $T$ dependence of $1/T_{1}$ cannot be exactly reproduced by any simple SC gap model, either with line nodes or without nodes, which may relate to the characteristics of the multiband SC state observed in Ba$_{0.6}$K$_{0.4}$Fe$_2$As$_2$\cite{ARPES}. 
%may depend on whether a system is electron-doped (LaFeAsO$_{0.7}$) or hole-doped (Ba$_{0.6}$K$_{0.4}$Fe$_2$As$_2$). 

These unconventional features of $1/T_1$ below $T_c$ were commonly observed in most FeAs-based superconductors \cite{Mukuda,Terasaki,Grafe,Nakai,Matano,Kotegawa,FukazawaSC}. In contrast, a fully-gapped SC state was observed in experiments such as ARPES \cite{ARPES} and magnetic penetration depth \cite{Penetrationdepth}. 
To reconcile these issues, $1/T_1$ was theoretically calculated on the basis of a nodeless extended $s_\pm$-wave pairing model with a sign reversal of the order parameter between the hole and electron Fermi surfaces \cite{Mazin,Kuroki}.  In the framework of either a two-band model, where the unitary scattering due to impurities is assumed \cite{NMRtheory}, or a five-band model in a rather clean limit \cite{NMRtheory2}, the experiments are well reproduced by such calculations. In fact, the results of $(1/T_1)$s for $^{57}$Fe and $^{75}$As in the SC state are consistent with the latter model. This may be because the intrinsic behavior of $1/T_1$ is measured for a highly homogeneous sample, which is guaranteed by the very sharp NMR linewidth. In this context, our results are consistently argued in terms of the extended s$_{\pm}$-wave pairing with a sign reversal of the order parameter among Fermi surfaces.  
Further, it would be desirable to measure $1/T_1$ at temperatures lower than 4 K and to systematically examine an impurity effect in these compounds.

\subsection{Normal-state properties of LaFeAsO$_{1-y}$ and Ba$_{0.6}$K$_{0.4}$Fe$_2$As$_2$}

\begin{figure}[tbp]
\begin{center}
\includegraphics[width=0.9\linewidth]{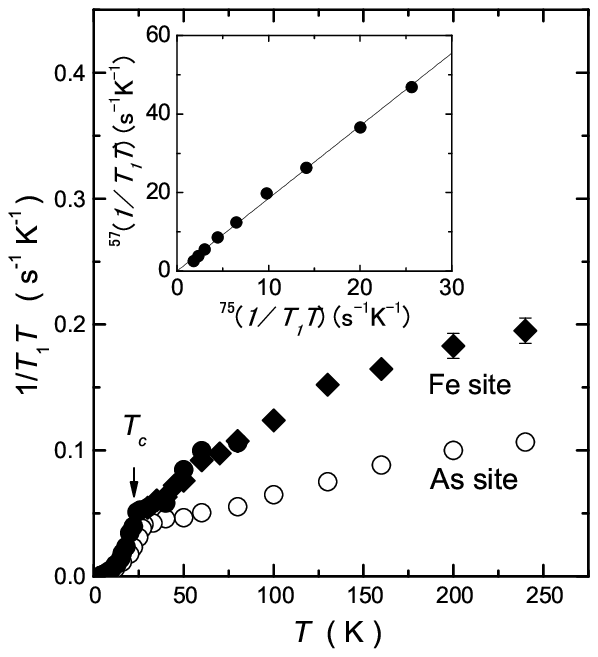}
\end{center}
\caption[]{ $T$ dependence of $^{57}$Fe-$1/T_{1}T$ at $H$ = 6.309 T ($\bullet$) and $H=$ 11.97 T (closed diamond), along with $^{75}$As-NQR-$1/T_{1}T$ in LaFeAsO$_{0.6}$($\circ$) \cite{Mukuda}. The inset shows the plot of $^{57}(1/T_{1})$ vs. $^{75}(1/T_{1})_{NQR}$  as the implicit parameter of $T$ between 30 and 240 K.
}
\label{fig:T1T}
\end{figure}

Next, we address the normal-state properties of LaFeAsO$_{1-y}$ and Ba$_{0.6}$K$_{0.4}$Fe$_2$As$_2$ through the $(1/T_1T)$ results. As shown in Fig. \ref{fig:T1T}, $^{57}(1/T_{1}T)$ at the Fe site for LaFeAsO$_{0.7}$ gradually decreases upon cooling down to $T_{c}$, resembling the behavior of $^{75}(1/T_{1}T)$ measured by NQR at the As site.  Actually, $^{57}(1/T_{1}T)$ at the Fe site is well scaled to $^{75}(1/T_{1}T)_{NQR}$ at the As site down to 60 K with a ratio of $^{57}(1/T_{1}T)/^{75}(1/T_{1}T)_{NQR}\simeq 1.85$.  It has been theoretically proposed that the multiple spin-fluctuation modes with {\bf Q} = ($\pi/a$, 0) and (0,$\pi/a$) originating from the nesting across the disconnected Fermi surfaces would mediate the extended s$_{\pm}$-wave pairing with a sign reversal of the order parameter. However, in our simple analyses of $(1/T_1T)$ results \cite{Terasaki}, we could only state that the spin fluctuations at finite wave vectors are more significant than the ferromagnetic spin fluctuation mode in this compound. Nevertheless, it is noteworthy that the $(1/T_1T)$s for both the Fe and As sites decrease upon cooling, indicating a  decrease in the low-energy spectral weight of spin fluctuations over the entire  $q$ space from room temperature. In contrast, in the case of the copper-oxide superconductors, $(1/T_1T)$s of $^{63}$Cu and $^{17}$O exhibit a different $T$ dependence due to the difference in the $q$-dependence of $^{63,17}A_{hf}(q)$ and the development of AFM spin fluctuations around $\textbf{Q}=(\pi/a,\pi/a)$ \cite{Takigawa}. The suppression of spin fluctuations over the entire $q$ space upon cooling below room temperature was observed in FeAs-based  superconductors, which has never been observed for other strongly correlated superconductors where an AFM interaction plays a vital role in mediating the Cooper pairing. 

\begin{figure}[htbp]
\begin{center}
\includegraphics[width=1\linewidth]{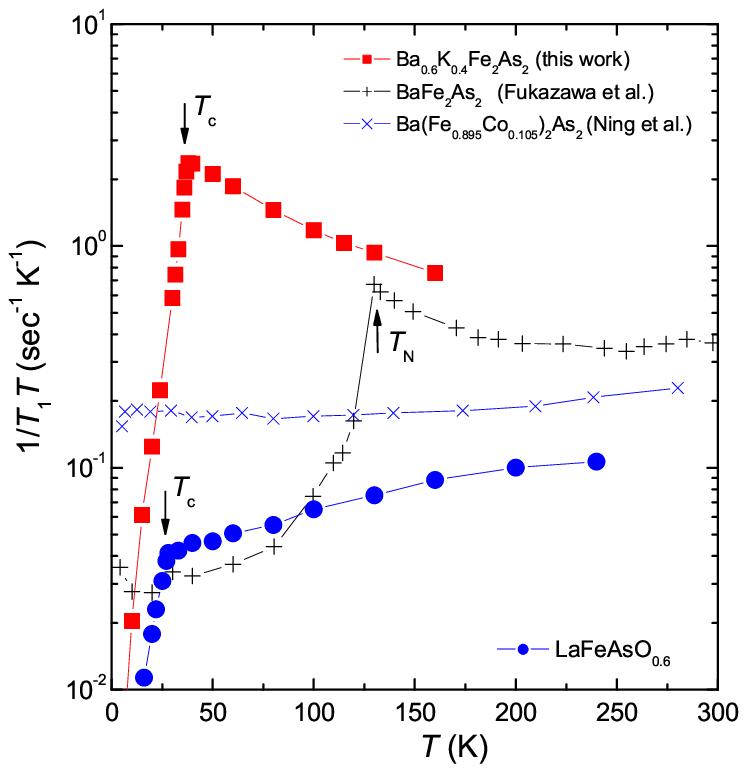}
\end{center}
\caption[]{(color online)
$T$ dependence of $1/T_{1}T$ for hole-doped Ba$_{0.6}$K$_{0.4}$Fe$_2$As$_2$ by $^{75}$As-NMR and for electron-doped LaFeAsO$_{0.6}$ by $^{75}$As-NQR, along with the $^{75}$As-NMR results for the undoped BaFe$_2$As$_2$ (cited from Fukazawa et al. \cite{Fukazawa}) and the electron-doped Ba(Fe$_{0.895}$Co$_{0.105}$)$_2$As$_2$ (cited from Ning et al. \cite{Ning}).
}
\label{fig:Ba122invT1T}
\end{figure}

Figure \ref{fig:Ba122invT1T} shows the $T$ dependence of $1/T_1T$ in the normal state of Ba$_{0.6}$K$_{0.4}$Fe$_2$As$_2$, along with the results of $^{75}$As-NQR $1/T_1$ for the electron-doped LaFeAsO$_{0.6}$ and $^{75}$As-NMR $1/T_1$s for the undoped BaFe$_2$As$_2$ \cite{Fukazawa} and the electron-doped Ba(Fe$_{0.895}$Co$_{0.105}$)$_2$As$_2$ \cite{Ning}.  It gradually increases upon cooling down to $T_c(H)$ = 37 K, in contrast to a significant decrease in the case of LaFeAsO$_{0.7}$. 
However, it should be noted that $1/T_1T$ in the electron-doped Ba(Fe$_{1-x}$Co$_{x}$)$_2$As$_2$ gradually decreases upon cooling and remains almost constant down to $T_c=20$ K below $\sim$100 K \cite{Ning}. It is remarkable that the $T$ dependence of $1/T_{1}T$ in the hole-doped Ba$_{0.6}$K$_{0.4}$Fe$_2$As$_2$ is significantly different from those in the electron-doped compounds such as Ba(Fe$_{1-x}$Co$_{x}$)$_2$As$_2$ and LaFeAsO$_{1-y}$; nevertheless, the SC characteristics  possess common features in these compounds.  Recently, on the basis of the fluctuation exchange approximation (FLEX) on an effective five-band Hubbard model, Ikeda found that with decreasing temperatures, $1/T_1T$ is enhanced in undoped and hole-doped systems. On the other hand, in electron-doped systems, it decreases significantly upon cooling, exhibiting a pseudogap behavior that originates from the band structure effect, that is, the existence of a high density of states just below the Fermi level. The effect becomes more remarkable with electron-doping. This qualitatively explains the NMR results. Such a pseudogap behavior exists even without electron correlation in the present band structure. Currently, the mechanism responsible for a pairing glue causing a possible extended s$_{\pm}$-wave pairing remains unknown.

\subsection{$^{75}$As-NQR studies of REFeAsO$_{1-y}$ (RE = La, Pr, Nd)}

\begin{figure}[htbp]
\begin{center}
\includegraphics[width=0.95\linewidth]{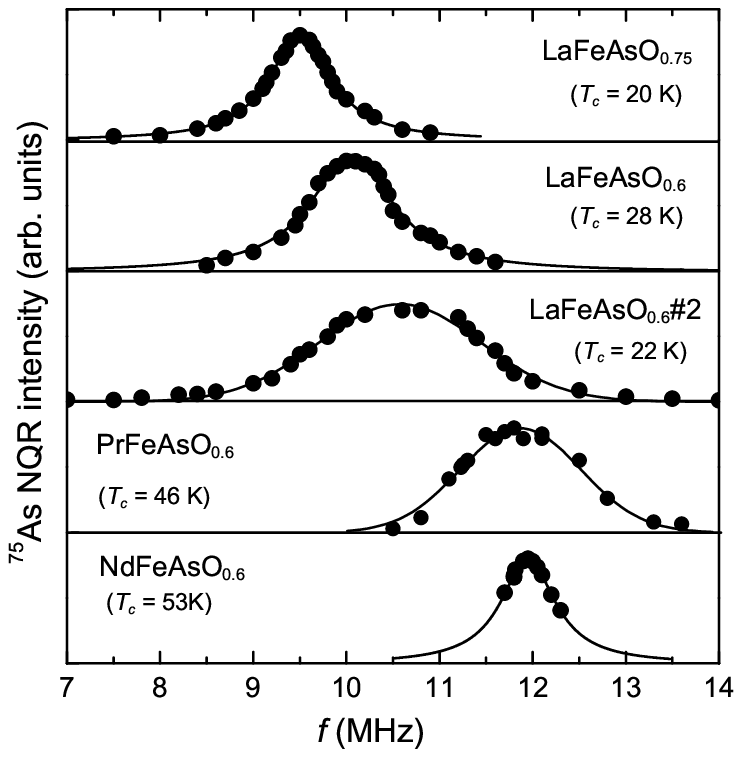}
\end{center}
\caption[]{$^{75}$As NQR spectra just above their $T_c$s for LaFeAsO$_{0.75}$ ($T_c=$~20 K), LaFeAsO$_{0.6}$ ($T_c=$~28 K), LaFeAsO$_{0.6}$($\sharp2$) ($T_c=$ 22 K), PrFeAsO$_{0.6}$ ($T_c=$ 46 K), and NdFeAsO$_{0.6}$ ($T_c=$ 53 K). }
\label{fig:AsNQR}
\end{figure}

\begin{figure}[htbp]
\begin{center}
\includegraphics[width=0.8\linewidth]{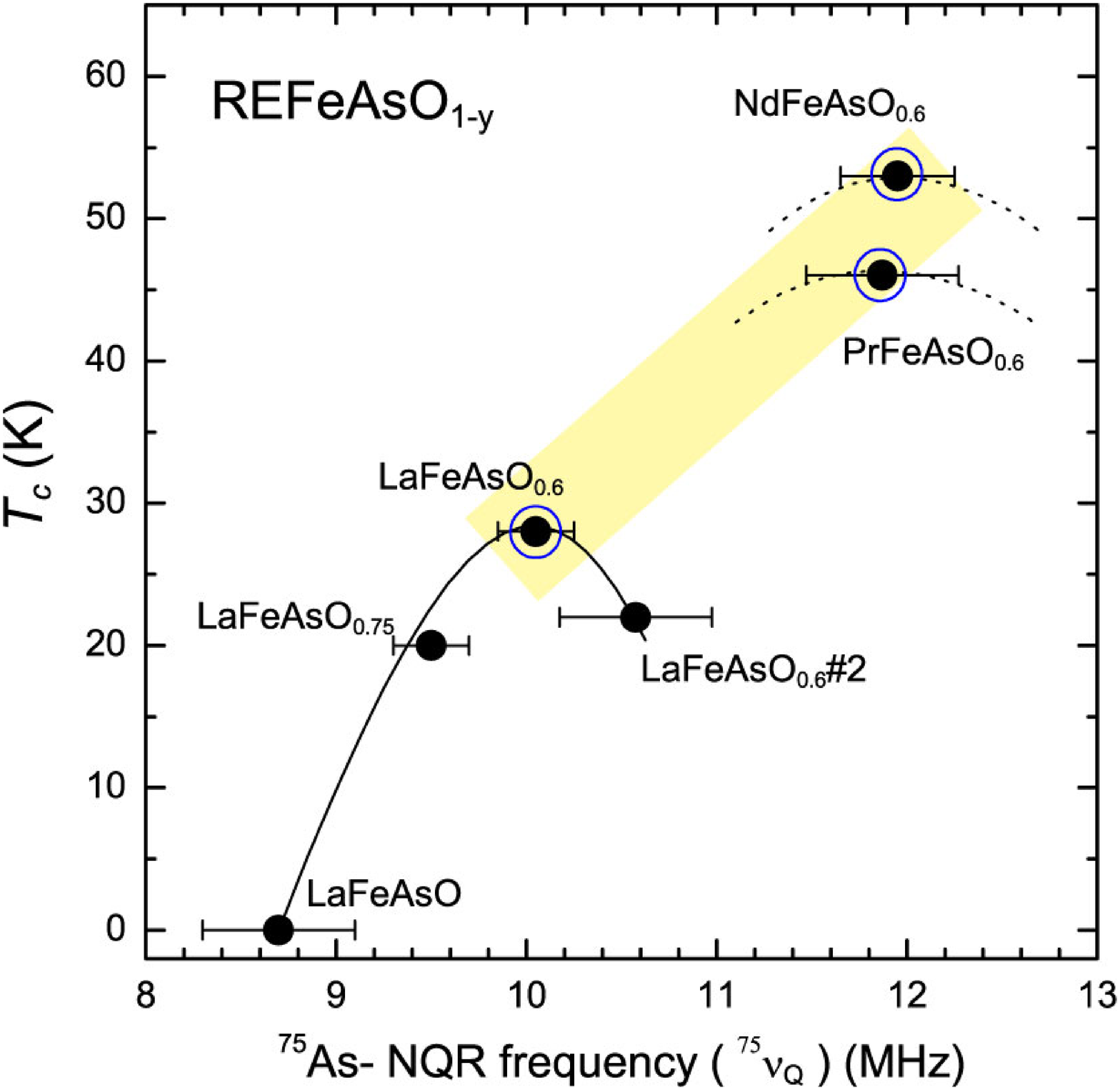}
\end{center}
\caption[]{(color online) A plot of $^{75}$As NQR frequency ($^{75}\nu_Q$) versus $T_c$ for LaFeAsO, LaFeAsO$_{0.75}$, LaFeAsO$_{0.6}$, PrFeAsO$_{0.6}$, and NdFeAsO$_{0.6}$. We have found an intimate relationship between the nuclear quadrupole frequency $\nu_Q$ of $^{75}$As and $T_c$ for the samples used in this study, revealing that $T_c$ is sensitive to the local configuration of the FeAs$_{4}$ tetrahedron.
}
\label{fig:NQR-Tc}
\end{figure}

Figure \ref{fig:AsNQR} shows the $^{75}$As-NQR spectra just above their $T_c$s for LaFeAsO$_{0.75}$ ($T_c=$~20 K), LaFeAsO$_{0.6}$ ($T_c=$~28 K), LaFeAsO$_{0.6}$($\sharp2$) ($T_c=$~22 K), PrFeAsO$_{0.6}$ ($T_c=$~46 K), and NdFeAsO$_{0.6}$ ($T_c=$ 53 K). 
A $^{75}$As-NQR frequency ($^{75}\nu_Q$) is obtained from the frequency at the peak of their $^{75}$As NQR spectra. 
Figure \ref{fig:NQR-Tc} shows a plot of  $T_c$ versus $^{75}\nu_Q$ for REFeAsO$_{1-y}$ (RE = La, Pr, Nd).  In the case of the LaFeAsO$_{1-y}$ system, this plot appears to exhibit a dome-like shape, having a maximum $T_c=28$ K at $^{75}\nu_Q=$ 10 MHz. Note that the respective $T_c$s = 46 K and 53 K of the optimally doped samples of REFeAsO$_{0.6}$ (RE = Pr and Nd) become larger than $T_c=28$ K of LaFeAsO$_{0.6}$ as $^{75}\nu_Q$ increases from 10 MHz in LaFeAsO$_{0.6}$ to 12 MHz in PrFeAsO$_{0.6}$ and NdFeAsO$_{0.6}$. This correlation between $T_c$ and $^{75}\nu_Q$ suggests an intimate relationship between the maximum value of $T_c$ and an optimum local structure, as revealed in literature \cite{C.H.Lee}.

$^{75}\nu_Q$ is proportional to the electric field gradient (EFG) along the c-axis $V_{zz}$. Here $\nu_Q=eQV_{zz}/2\sqrt{1+\eta^2/3}$, where $Q$ is the nuclear quadrupole moment of $^{75}$As and $\eta$ is the asymmetry parameter of the EFG. The EFG is generally given by two contributions; one is a non-cubic charge distribution of $4p$ orbitals at the $^{75}$As site and the other is the charge distribution arising from the surrounding ions around the $^{75}$As site, denoted by $V_{zz}^{\rm in}$ and $V_{zz}^{\rm out}$, respectively. The former originates from the hybridization between the As-$4p$ orbitals and Fe-$3d$ orbitals in the FeAs layer, and the latter may have a predominant contribution relevant to the Madelung potential originating from the charge distributions of the neighboring Fe atoms and REO$_{1-y}$ layers. The variation of lattice parameters through doping significantly influences $V_{zz}^{\rm out}$, resulting in a dome-like shape in the plot of $T_c$ versus $^{75}\nu_Q$ for LaFeAsO$_{1-y}$, as shown in Fig. \ref{fig:NQR-Tc}. 
In fact, the lengths of the a- and c-axes in the tetragonal structure decrease with oxygen content in LaFeAsO$_{1-y}$, and it decreases upon the replacement of La with Nd in REFeAsO$_{0.6}$. 
Despite the reduction in the lattice volume, the neutron diffraction experiment by Lee {\it et al.} has revealed that the distance between the Fe- and As-planes becomes larger in NdFeAsO$_{0.6}$ than in LaFeAsO$_{0.6}$ \cite{C.H.Lee}. By assuming the point charges of the surrounding ions around the $^{75}$As site, a simple calculation of $V_{zz}^{\rm out}$ has revealed that the $V_{zz}^{\rm out}$ becomes larger for NdFeAsO$_{0.6}$ than for LaFeAsO$_{0.6}$ and LaFeAsO. 
However, the calculated values of $^{75}\nu_Q$ cannot reproduce the experiments quantitatively, indicating that the on-site contribution  $V_{zz}^{\rm in}$ is also important in these compounds. Namely, the change in the  distance between the Fe- and As-plane varies the charge distribution of the As-$4p$ orbitals, increasing $^{75}\nu_Q$ in going from non-superconducting LaFeAsO to NdFeAsO$_{0.6}$ with $T_c=$ 53 K.   
The variation of the hybridization between As-$4p$ orbitals and Fe-$3d$ orbitals induces the modification of the Fe-As layer-derived band structure as well.  Here, we note that $^{75}\nu_Q=2.2$ MHz for undoped BaFe$_2$As$_2$ is significantly lower than the value of 8.7 MHz for undoped LaFeAsO. The variation of lattice parameters due to the change in the crystal structure is expected to mainly influence $V_{zz}^{\rm out}$.
Interestingly, $^{75}\nu_Q$ in the BaFe$_2$As$_2$ system increases with K doping, suggesting that an increase in either the carrier density or the hybridization between As-$4p$ orbitals and Fe-$3d$ orbitals leads to an increase in $^{75}\nu_Q$. 

Therefore, the intimate relationship between $\nu_Q$ and $T_c$ in REFeAsO$_{1-y}$ suggests that the local configuration of Fe and As atoms is significantly related to the $T_c$ of Fe-based pnictide superconductors, that is, $T_c$ can be enhanced up to 50 K when the local configuration of Fe and As atoms becomes optimal. Here, it may be relevant that $T_c$ becomes maximum when the bonding angle between Fe-As-Fe coincides with that of a regular tetrahedron of FeAs$_4$ \cite{C.H.Lee}.

%%%%%%%%%%%%%%%%%%%% Summary %%%%%%%%%%%%%%%%%%%

\section{Summary}

$^{57}$Fe-NMR and $^{75}$As-NQR/NMR studies have clarified the novel SC and normal-state characteristics of $^{57}$Fe-enriched LaFeAsO$_{0.7}$ ($T_{c}$ = 28 K) and Ba$_{0.6}$K$_{0.4}$Fe$_2$As$_2$ ($T_c$ = 38 K). 
In the SC state of LaFeAsO$_{0.7}$ ($T_{c} =$ 28 K), the spin component of the  $^{57}$Fe-Knight shift decreases to almost zero at low temperatures, which provide firm evidence of a superconducting state formed by spin-singlet Cooper pairing. 
The measurements of the Knight shift and $T_1$ have revealed that an extended s$_{\pm}$-wave pairing with a sign reversal of the order parameter can be a promising candidate.

In the normal state of LaFeAsO$_{0.7}$, we have found a remarkable decrease in $1/T_1T$ upon cooling for both the Fe and As sites, whereas $1/T_1T$ gradually increases upon cooling down to $T_c$ in the case of Ba$_{0.6}$K$_{0.4}$Fe$_2$As$_2$. Remarkably, the $T$ dependence of $1/T_{1}T$ in the normal state drastically changes when going from the hole-doped compound Ba$_{0.6}$K$_{0.4}$Fe$_2$As$_2$ to electron-doped compounds such as Ba(Fe$_{1-x}$Co$_{x}$)$_2$As$_2$ and LaFeAsO$_{1-y}$; nevertheless, the SC characteristics are not drastically different among these compounds. 

Recently, on the basis of the fluctuation exchange approximation (FLEX) on an effective five-band Hubbard model, Ikeda found that with decreasing temperatures, $1/T_1T$ in an electron-doped system decreases significantly upon cooling, exhibiting a pseudogap behavior that originates from the band structure effect, that is, the existence of a high density of states just below the Fermi level\cite{Ikeda}. This qualitatively explains the NMR results. Such pseudogap behavior exists even without electron correlation in the present band structure. 
%In this context, it is not always concluded that AFM spin fluctuations are responsible for the enhancement of $1/T_1T$ upon cooling in  hole-doped Ba$_{0.6}$K$_{0.4}$Fe$_2$As$_2$. 
Currently, the mechanism  responsible for a pairing glue causing a possible extended s$_{\pm}$-wave pairing remains unknown.  Further experiments on the $T_c$ dependences of $1/T_1$ and $K$ at both the Fe and As sites using a single crystal are required to understand the nature of spin fluctuations.

\section*{Acknowledgements}

We are grateful to N. Tamura, H. Yamashita, and H. Kinouchi for their assistance with some parts of NMR/NQR measurements. We are grateful to K. Miyazawa, P.M. Shirage, H. Kito, K. Kihou, and H. Eisaki for providing the crystals of REFeAsO$_{1-y}$ and (Ba,K)Fe$_2$As$_2$, and S. Suzuki, S. Miyasaka, and S. Tajima for providing LaFeAsO.
This work was supported by a Grant-in-Aid for Specially Promoted Research (20001004) and by the Global COE Program (Core Research and Engineering of Advanced Materials-Interdisciplinary Education Center for Materials Science) from the Ministry of Education, Culture, Sports, Science and Technology (MEXT), Japan.

%---------------------------- Fig.1 NMR spectrum ---------------------
%----------------------------------------------------------------------
%---------------------------- Fig.2 NMR spectrum ---------------------
%----------------------------------------------------------------------

%--------------------  Fig.3 Recovery curve  ---------------------------
%----------------------------------------------------------------------

%------------------- Fig.4 T dependence of 1/T1 ----------------------
%----------------------------------------------------------------------

%--------------------  Fig.5 Recovery curve  ---------------------------
%----------------------------------------------------------------------
%**************  Fig.6 *************************************************
%****************************************************************************
%**************  Fig.7  **************************************************
%****************************************************************************
%**************  Fig.8  *************************************************
%****************************************************************************
%-------------------- Fig.9 -------------------------
%----------------------------------------------------------------------
%**************  Fig.10  ********************************************
%****************************************************************************
%**************  Fig. 11**************************************************
%****************************************************************************

%**************  Fig. 12  *********************************************
%****************************************************************************

\end{document}